\documentclass[11pt]{article}
\usepackage{latexsym}
\usepackage{graphics}
\usepackage{amsmath}
\usepackage{amsfonts}
\newcommand{\bmat}{\left(\begin{array}}
\newcommand{\emat}{\end{array}\right)}
\newcommand{\beq}{\begin{equation}}
\newcommand{\eeq}{\end{equation}}
\newcommand{\sub}[1]{\phantom{}_{(#1)}\phantom{}}

\def\yzero{\smash{\hbox{$y\kern-4pt\raise1pt\hbox{${}^\circ$}$}}}
\def\p{\partial}
\def\a{\alpha}
\def\b{\beta}
\def\g{\gamma}
\def\d{\delta}

\def\-{\hphantom{-}}

\def\s2{\frac{1}{\sqrt2}}

\def\beq{\begin{equation}}
\def\eeq{\end{equation}}
\def\beqa{\begin{eqnarray}}
\def\eeqa{\end{eqnarray}}

\def\Tr{{\rm Tr \,}}

\def\IF{\relax{\rm I\kern-.18em F}}
\def\II{\relax{\rm I\kern-.18em I}}
\def\IP{\relax{\rm I\kern-.18em P}}

\def\Dsl{\,\raise.15ex\hbox{/}\mkern-13.5mu D} 

\def\IC{\bf C}
\def\IZ{\bf Z}

\def\z2z2{$\IC^3/(\IZ_2\times\IZ_2)$}

\def\a{\alpha}
\def\b{\beta}

\def\d{\delta}
\def\e{\epsilon}
\def\f{\phi}
\def\g{\gamma}

\def\k{\kappa}
\def\l{\lambda}
\def\m{\mu}
\def\n{\nu}

\def\p{\pi}

\def\r{\rho}
\def\s{\sigma}

\def\z{\zeta}
\def\D{\Delta}
\def\F{\Phi}
\def\G{\Gamma}

\def\L{\Lambda}

\def\S{\Sigma}

\def\ca{{\cal A}}

\def\cl{{\cal L}}

\def\co{{\cal O}}

\def\bo{{\raise-.3ex\hbox{\large$\Box$}}}      
\def\pa{\partial}                              
\def\face{{\raise.2ex\hbox{$\displaystyle \bigodot$}\mskip-2.2mu \llap {$\ddot
        \smile$}}}                              

\def\leftrightarrowfill{$\mathsurround=0pt \mathord\leftarrow \mkern-6mu
        \cleaders\hbox{$\mkern-2mu \mathord- \mkern-2mu$}\hfill
        \mkern-6mu \mathord\rightarrow$}       
\def\dvec#1{\vbox{\ialign{##\crcr
        \leftrightarrowfill\crcr\noalign{\kern-1pt\nointerlineskip}
        $\hfil\displaystyle{#1}\hfil$\crcr}}}           

\def\beq{\begin{equation}}
\def\eeq{\end{equation}}
\def\beqx{\begin{displaymath}}
\def\eeqx{\end{displaymath}}
\def\beqa{\begin{eqnarray}}
\def\eeqa{\end{eqnarray}}
\def\NO{\nonumber}

\def\be{\begin{equation}}
\def\ee{\end{equation}}
\def\bea{\begin{eqnarray}}
\def\eea{\end{eqnarray}}

\def\bop#1{\setbox0=\hbox{$#1M$}\mkern1.5mu
        \vbox{\hrule height0pt depth.04\ht0
        \hbox{\vrule width.04\ht0 height.9\ht0 \kern.9\ht0
        \vrule width.04\ht0}\hrule height.04\ht0}\mkern1.5mu}
\def\Box{{\mathpalette\bop{}}}                        
\def\pa{\partial} 

\def\<{\langle}
\def\>{\rangle}

\begin{document}

\begin{flushright}
ITFA-2004-17 \\
{\tt hep-th/0404176}
\end{flushright}

\vskip 2cm

\begin{center}
{\Large \bf  AdS/CFT correspondence and Geometry \\}
\vskip 1cm

{\bf Ioannis Papadimitriou\footnote{ipapadim@science.uva.nl} and 
Kostas Skenderis\footnote{skenderi@science.uva.nl}} 
\vskip 0.5cm

{\it Institute for  Theoretical Physics, University of Amsterdam,\\
 Valkenierstraat 65, 1018 XE Amsterdam, The Netherlands.} \\

\end{center}

\vskip1cm

\begin{center}
{\bf Abstract} 
\end{center}
\medskip

In the first part of this paper we provide a short introduction 
to the AdS/CFT correspondence and to holographic renormalization.
We discuss how QFT correlation functions, Ward identities and anomalies
are encoded in the bulk geometry. In the second part 
we develop a Hamiltonian approach to the method of holographic
renormalization, with the radial coordinate
playing the role of time. In this approach regularized
correlation functions are related to canonical momenta
and the near-boundary expansions
of the standard approach are replaced by covariant expansions
where the various terms are organized according to their dilatation 
weight. This leads to universal expressions for counterterms and
one-point functions (in the presence of sources)  
that are valid in all dimensions.
The new approach combines optimally elements from all 
previous methods and supersedes them in efficiency.

\newpage

\tableofcontents

\section{Introduction}
   
The AdS/CFT correspondence \cite{Maldacena:1997re,Gubser:1998bc,Witten:1998qj}
(for reviews see \cite{Aharony:1999ti,D'Hoker:2002aw}) relates string theory 
on (locally asymptotically) AdS spacetimes (times a compact space) with 
a quantum field theory (QFT) ``residing'' on the conformal boundary 
of the bulk spacetime.\footnote{In the first examples
discussed in the literature, the bulk spacetime
was exactly AdS (times a compact space) and the dual theory 
was a conformal field theory (CFT). This motivated the name 
``AdS/CFT correspondence''. Our discussion is applicable under 
the more general circumstances mentioned above.} 
In a specific limit, which is a strong 
coupling limit of the boundary theory, the bulk theory 
reduces to classical gravity coupled to certain matter fields.
In this limit QFT data is encoded in classical geometry.
The aim of this contribution is to discuss how to extract 
this data from the bulk geometry. 

In the first part of this paper we give a brief review of the AdS/CFT
correspondence and of holographic renormalization for non-experts.  
We start our discussion from the QFT side by reviewing 
what QFT data we would like to obtain from gravity.
This data consists of correlation functions of 
gauge invariant operators and of symmetry relations.
Symmetries of the QFT action imply relations among correlation functions,
the so-called Ward identities. Such relations are ``kinematical''
and can be established without the need to actually compute the 
correlation functions. Sometimes, however,
quantum effects imply that some of the classical symmetries
are broken at the quantum level: the corresponding Ward identities 
are anomalous. The symmetries of the quantum theory are thus
encoded in the corresponding Ward identities and anomalies.

The next task is to describe how to obtain correlation
functions, Ward identities and anomalies using the AdS/CFT correspondence. 
We outline the prescription of \cite{Gubser:1998bc,Witten:1998qj}
for the computation of correlation functions and 
discuss how to deal with the infinities arising 
in such computations. This is dealt with via the 
formalism of holographic renormalization \cite{HS,dHSS,BFS} 
(for a review see \cite{Skenderis:2002wp}; for related work 
see \cite{Balasubramanian:1999re,Kraus:1999di} -- a more complete 
list of references can be found in the review).
This formalism automatically incorporates the ``kinematical''
constraints, i.e. the Ward identities
and their anomalies, and identifies the part of the 
geometry where the ``dynamical'' information, i.e. the 
correlation functions, is encoded.  A central role in this method 
is played by the fact that one can perturbatively 
work out the asymptotics of all bulk fields to sufficiently
high order using the radial distance
from the boundary of AdS as a small parameter\footnote{
From the point of view of the dual field theory we expand around a UV fixed
point, the small parameter being the inverse energy.} \cite{FeffermanGraham}
(for relevant math reviews see \cite{Graham,anderson}). Correlation
functions are encoded in specific coefficients in the 
asymptotic expansion of the bulk fields and Ward identities
and anomalies originate from certain relations that 
these coefficients satisfy. 

This method, even though complete, is not very efficient as 
we discuss later. The last part of the paper is devoted to 
developing a ``Hamiltonian'' version of the method, where 
the radial coordinate plays the role of time.
In this approach the focus is shifted from the on-shell
supergravity action to the canonical momenta of the bulk fields.
The latter are associated with (regularized) correlation 
functions of gauge invariant operators \cite{deBoer:1999xf}. 
To obtain renormalized correlation functions we need 
to subtract the infinities. This was done in the standard approach 
via the near-boundary analysis. In the new approach we 
use instead the fact that there is a well defined dilatation operator. 
This allows us to develop a covariant expansion of the asymptotic solution 
where the various terms are organized according to their dilatation
eigenvalue.  This leads to a faster algorithm for determining counterterms 
and correlation functions.\footnote{A different Hamiltonian approach 
to renormalization using 
the Hamilton-Jacobi equation \cite{deBoer:1999xf}
was developed in \cite{Martelli:2002sp}, see also 
\cite{Fukuma:2002sb} and references therein.}
In particular, we obtain universal recursion 
relations for the asymptotic solutions and counterterms
that are valid in all dimensions. 

This paper is organized as follows. In the next section we 
discuss the QFT data that enters in the discussion of the 
AdS/CFT correspondence. The discussion is illustrated by 
a number of simple examples and assumes only a general familiarity with
QFT. In section 3 we present a brief review of the AdS/CFT correspondence and 
of holographic renormalization. More details can be found in the 
reviews listed above. Sections 2 and 3 are aimed at non-experts 
that want to get a flavor of the ideas and techniques involved 
in the AdS/CFT correspondence.  Experts can safely move directly 
to section 4 where the new Hamiltonian formulation is 
discussed systematically. This section is self-contained and can be 
read independently of the previous sections.

Throughout this paper we work with Euclidean signature. All 
results, however, can be straightforwardly continued to
any other signature.

\section{QFT data}

We discuss in this section the quantum field theory data that 
we would like to extract from the bulk geometry. 
We would say that a QFT is solved if we determine all correlation
functions of all gauge invariant operators. The set of gauge 
invariant operators depends on the theory under consideration.
Examples of such operators are the stress energy tensor $T_{ij}$, 
currents $J^i$ associated with global symmetries 
and scalar operators $O$. As mentioned in the introduction, 
when the bulk spacetime is exactly $AdS$ (times a compact space) 
the dual quantum field theory is a conformal field
theory (CFT). The discussion in this section will refer to CFTs, 
but all considerations have a straightforward generalization 
to QFTs that can be viewed either as deformations of the 
CFTs by relevant or marginal operators or to CFTs with 
spontaneously broken conformal invariance.
 
The examples of AdS/CFT correspondence involve specific 
CFTs -- the most studied case being the maximally supersymmetric 
gauge theory in four dimensions, the  $\mathcal{N}=4$ SYM theory.
The discussion below is focused on general properties that 
do not depend on the details of the specific CFT. 
Given a (perturbative) CFT specified by set of fields $\varphi^A$
one can work out the set of gauge invariant composite 
operators $\co(\varphi^A)$. Their correlation
functions can then be computed in perturbation theory.
To give an elementary example
of a CFT correlator consider a scalar operator $O_\D$ of conformal weight $\D$.
In this case the form of the 2-point function is fixed by conformal
invariance, 
\be \label{2pt}
\langle O_\D(x) O_\D(0) \rangle = \frac{c(g,\D)}{x^{2 \D}},
\ee
where $c(g,\D)$ is a constant that depends on the coupling constant
of the theory $g$ and the conformal dimension $\D$ of the operator. One may 
set it to one by a choice of normalization of $O_\D$ but we shall not 
do so. Our objective is to understand how to extract this 
and more complicated higher point functions (that are not determined 
by symmetries) from the bulk geometry. 
  
In general, symmetries 
of the classical action imply relations among correlation 
functions, the so-called Ward identities. To give an example:
Poincar\'{e} invariance of the classical action 
implies (classically) that the stress energy tensor 
is conserved,
\be
\pa^i T_{i j}=0.
\ee
At the quantum level this implies relations among certain correlation 
functions. For instance,
\be \label{2ptd}
\pa_x^i \langle T_{i j}(x) O(y) O(z) \rangle
= \pa_j \d(x-y) \langle O(x) O(z) \rangle 
+ \pa_j \d(x-z) \langle O(y) O(x) \rangle. 
\ee

Some of the classical symmetries, however, are broken by quantum 
effects. For example, the stress energy tensor of a field theory that 
is classically conformally invariant is traceless, but quantum 
effects may break this symmetry
\be \label{tmm}
T^i_i =0 \quad classical, 
\qquad \qquad \langle T^i_i\rangle =\ca \quad quantum.
\ee
Since $T^i_i$ generates scale transformations, the conformal anomaly 
captures the fact that the correlators, even though they
are CFT correlators, are not scale invariant,
\be \label{WWI}
\m \frac{\d}{\d \m} \langle O_1(x_1) \cdots O_n(x_n) \rangle 
= A \d(x_1,\dots,x_n)
\ee 
where $\d(x_1,\dots,x_n)=\d(x_1-x_2) \d(x_2-x_3) ... \d(x_{n-1}-x_n)$,
and $A$ is related to $\ca$ in a way we specify below.
Notice that the violation of conformal invariance is a contact term.
In a general QFT (i.e. not conformal) (\ref{WWI}) is replaced by the 
beta function equation.

To understand how an anomaly can arise consider the 2-point function
in (\ref{2pt}). The form of this correlator (for  $x^2 \neq 0$) is completely 
fixed (up to normalization) by conformal invariance. 
Depending on the conformal dimension, however, the correlator
may suffer from short distance singularities. Consider the case
$\D \sim d/2 + k$, where $d$ is the spacetime dimension and 
$k$ is an integer. As $x^2 \to 0$ the correlator behaves as
\be
\frac{1}{x^{2 \D}} \sim \frac{1}{d + 2 (k - \D)} 
\frac{\G(d/2)}{2^{2 k} k! \G(d/2 +k)} S^{d-1} \Box^k \d^{(d)}(x)
\ee
where $S^{d-1}=2 \p^{d/2}/\G(d/2)$ is the volume of the unit 
$(d-1)$-sphere
and $\Box=\d^{ij} \pa_i \pa_j$. We thus find that there
is a pole at $\D = d/2 + k$.
To produce a well-defined distribution we use 
dimensional regularization and subtract the pole. For concreteness we 
consider the case $k=0$ (all 
other cases follow by differentiation w.r.t. to $x$, see \cite{PS}).
Minimal subtraction yields \cite{OP,PS}
\bea
\langle O_{d/2}(x) O_{d/2}(0) \rangle_R & = & c(g,d/2) \lim_{\D \to d/2} 
\left(\frac{1}{x^{2 \D}} 
- \frac{\mu^{2 \D -d}}{ d-2 \D} S^{d-1} \d^{(d)}(x)\right) \nonumber\\
&=& - c(g,d/2) \frac{1}{2(d-2)} \Box 
\frac{1}{(x^2)^{{\frac{1}{2}d}-1}}\left(\log \m^2 x^2 
+ \frac{2}{d-2}\right)\,.
\eea
where the subscript $R$ indicates that this is a renormalized correlator.
The scale $\mu$ is introduced, as usual in dimensional regularization,
on dimensional grounds. 
The renormalized correlator agrees with the bare one away 
from coincident points but is also well-defined at $x^2=0$.
Let us now consider the scale dependence of the renormalized correlator,
\be \label{d/2}
\m \frac{\pa}{\pa \m} \langle O_{d/2}(x) O_{d/2}(0) \rangle_R 
=S^{d-1} c(g,d/2)
\d^{(d)}(x).
\ee
where we used $\Box (x^{2})^{-d/2+1} = -(d-2) S^{d-1} \d^{(d)}(x)$.
Thus the renormalized correlation function exhibits a violation
of scale invariance. We shall soon connect this to the violation of the 
tracelessness condition of the stress energy tensor.

Recall that correlation function of composite operators may be computed 
by introducing sources that couple to them. The generating functional
of correlation functions then has the following path 
integral representation,
\be \label{Z}
Z[g_{(0)}, \f_{(0)}]= 
\int [D \varphi^A] 
\exp \left(-\int d^dx \sqrt{g_{(0)}}[\cl_{CFT}(\varphi^A;g_{(0)}) 
+ \f_{(0)} O(\varphi^A)]\right)
\ee
where $\varphi^A$ represents collectively all fields of the theory, 
$g_{(0)}$ is 
a background metric (which serves as a source for the stress energy tensor), 
$\cl_{CFT}$ is the Lagrangian density for the CFT 
and $\f_{(0)}$ is a source for the operator $O$. 
Correlation functions 
can now be computed by differentiating w.r.t. sources and then setting the sources to 
zero. For instance, the connected two-point function of $O$ on flat spacetime 
($g_{(0)ij}=\d_{ij}$)
is given by
\be
\langle O(x) O(0) \rangle = \left.
\frac{ \delta^2 W}{\d \f_{(0)}(x) 
\delta \f_{(0)}(0)}\right|_{\f_{(0)}=0}
\ee
where $W=\log Z$ is the generating functional of connected correlators.
Given a Lagrangian density $\cl_{QFT}$ one could thus compute the correlation
functions of $O$ by first (perturbatively) computing $Z[g_{(0)},\f_{(0)}]$.
Such computations however are plagued by infinities and to make sense of 
them one needs to renormalize the theory. To subtract the divergences
one may add counterterms to the action. If the counterterms break a
classical symmetry, then this symmetry is anomalous.

A slightly different route is to first compute in general the one-point 
functions in the presence of sources,
\be \label{1ptqft}
\langle T_{ij}(x) \rangle_s = -\frac{2}{\sqrt{g_{(0)}(x)}} 
\frac{ \delta W[g_{(0)},\f_{(0)}]}{\delta g^{ij}_{(0)}(x)}, \qquad
\langle O(x) \rangle_s =- \frac{1}{\sqrt{g_{(0)}(x)}} 
\frac{ \delta W[g_{(0)},\f_{(0)}]}{\delta \f_{(0)}(x)} 
\ee
where the subscript $s$ in the correlation functions indicate that the 
sources are non-zero. Correlation functions are then computed by further
differentiating w.r.t. sources and setting the sources to zero.
This reformulation will be proved useful later when we show how to
compute correlation functions holographically (i.e. using the AdS/CFT
correspondence). Another advantage is that 
one can express compactly many Ward identities. For instance, invariance
of $Z$ under diffeomorphisms, 
\be
\d g_{(0)}^{ij} = -(\nabla^i \xi^j + \nabla^j \xi^i), \qquad
\d \f_{(0)} = \xi^j \nabla_j \f_{(0)}
\ee
implies
\be
\nabla^i \langle T_{ij}(x) \rangle_s = - \< O(x) \>_s \nabla_j \f_{(0)}(x)  
\ee 
Differentiating now twice w.r.t. $\f_{(0)}$ and then setting 
$\f_{(0)}=0$, $g_{(0)ij}=\d_{ij}$ leads to (\ref{2ptd}).

Using the fact that the trace of the stress energy tensor is 
the generator of conformal transformations we arrive at \cite{PS}
\be \label{wsc}
\int d^d x \sqrt{g_{(0)}} g_{(0)}^{ij} \< T_{ij} \> 
= \sum_{k=1}^\infty \frac{(-1)^k}{k!} \int \prod_{i=1}^k \left(
d^d x_i \sqrt{g_{(0)}} J(x_i) \right) \mu \frac{\pa}{\pa \m} 
\< \co(x_1) \cdots \co(x_k) \>
\ee 
where $J$ denotes all sources and $\co$ the corresponding operators.
In our case, $J=\{\f_{(0)}, g_{(0)ij} \}$ and $\co=\{O, T_{ij}\}$.
Clearly, the expectation value of the stress energy tensor 
is non-vanishing if the scale derivative of the correlator is
non-vanishing. 
In particular, we have seen in (\ref{d/2}) that the scale derivative
of the 2-point function of an operator of dimension $d/2$ yields
a delta function. Inserting this in (\ref{wsc}) we obtain
\be
\< T^i_i\> = \frac{1}{2} S^{d-1} c(g,d/2) \f_{(0)}^2
\ee
So, in this case, $\mathcal{A}=A\f_{(0)}^2/2$ and $A= S^{d-1} c(g,d/2)$. 
This result 
generalizes to all operators of dimension $\D=d/2+k$
with result \cite{PS}
\be
\< T^i_i\> = \frac{1}{2} c_k \f_{(0)} \Box^k \f_{(0)}, \qquad
c_k = \frac{\pi^{d/2}}{2^{2 k-1} \Gamma(k+1) \G(k+d/2)} c(g,\D).
\ee
These considerations were valid for flat spacetime. When the 
background is curved, the results generalize to
\be \label{anomaly}
\< T^i_i\> = \frac{1}{2} c_k \f_{(0)} P_k \f_{(0)} + \left(a E +
\sum_i c_i W^i\right)  + \nabla_i J^i
\ee
$P_k$ is equal to  $\Box^k$ when the background is flat
and transforms covariantly under Weyl transforms
$g_{(0)} \to g_{(0)} e^{2 \s}$
\be
P_k \to e^{-(d/2+ k) \s} P_k e^{(d/2-k) \s}
\ee
For instance, for $k=1$,
\be
P_1 = \Box + \frac{d-2}{4(d-1)} R.
\ee
The two terms inside the parenthesis in (\ref{anomaly}) are purely gravitational
and are present only when $d$ is even. $E$ is the Euler density,
$W^i$ is a basis of Weyl invariants of dimension $d$ and $a$ and $c_i$
are numerical constants that depend on the field content of the theory. For instance, in $d=4$
there is one Weyl invariant (the square of the Weyl tensor),
in $d=6$ there are three such tensors, etc.
The last term in (\ref{anomaly}) is scheme dependent, i.e. it can be 
modified by
local finite counterterm terms in the action. In general there may be
additional terms in (\ref{anomaly}) that depend on higher powers of
the sources $\phi_{(0)}$. These would be related to singularities
in higher-point functions.
The structure of (\ref{anomaly}) is dictated by the fact that the
integrated conformal anomaly is itself conformally invariant 
\cite{Bonora:1985cq,Deser:yx}.

The AdS/CFT duality implies that all this data is encoded in the
geometry. We discuss in the next section how to recover them 
from the bulk geometry.

\section{AdS/CFT correspondence}

The AdS/CFT correspondence states that there is an
exact equivalence between string theory on (locally) asymptotically AdS (AAdS)
spacetimes (times a compact space) and a quantum field theory that
``resides'' on the conformal boundary of the AAdS spacetime. 
In the regime where the one description is perturbative the other
one is strongly coupled. We will work in the regime where the
gravitational description is valid and we will describe how to
obtain the QFT data described in the previous section.

The basic AdS/CFT dictionary is as follows: 
\begin{enumerate}
\item Gauge invariant operators of the boundary theory are in one-to-one 
correspondence with bulk fields. For example, the bulk metric corresponds 
to the stress energy tensor of the boundary theory. 
\item  The leading boundary behavior of the bulk field is 
identified with the source of the dual operator. 
\item The string partition function (which is a functional of the fields
parameterizing the boundary behavior of the bulk fields) is identified
with the generating functional of QFT correlation functions.
\end{enumerate}
At low energies and to leading order the AdS/CFT prescription reads,
\be
S_{{\rm on-shell}}[f_{(0)}]=-W[f_{(0)}]
\ee
where $S_{{\rm on-shell}}[f_{(0)}]$ is the on-shell value of the supergravity
action, $f_{(0)}$ denotes collectively all fields parameterizing the
boundary values of bulk fields
and $W$ is the generating functional of connected
graphs (see the discussion below (\ref{Z})).
It follows that one can compute correlators of the (strongly
coupled) QFT gravitationally by first evaluating the on-shell
value of the supergravity action and then differentiating w.r.t.
the boundary values, e.g. 
\be
\langle O(x) O(0) \rangle = \left.
-\frac{ \delta^2 S_{{\rm on-shell}}}{\d \f_{(0)}(x) 
\delta \f_{(0)}(0)}\right|_{\f_{(0)}=0}
\ee
A naive use of these formulas however yields infinite answers --
the on-shell value of the action is infinite due to the infinite
volume of the AAdS spacetime. The goal of holographic renormalization
is to make these formulas well-defined.

The general form of the bulk action is 
\be
S = \int d^{d+1} x \sqrt{g}[-\frac{1}{2 \k^2} R 
+ \frac{1}{2} g^{\m \n} \pa_\m \F \pa_\n \F + V(\Phi) + \cdots]
\ee
where $\k^2 = 8 \p G_{d+1}$ ($G_{d+1}$ is Newton's constant) and 
the dots indicate contribution of additional fields such as
gauge fields, fermions, antisymmetric tensors. The analysis below
generalizes straightforwardly to include such fields (but it becomes a lot more tedious). 
Restricting to the gravity-scalar sector
means that we only study correlation functions of the
stress energy tensor and a scalar operator. The potential has
the form,
\be
V(\F) = \frac{\L}{\k^2} +\frac12 m^2 \F^2 + g \F^3 + \cdots
\ee
where $\L$ is the cosmological constant and the mass $m^2$ of the
scalar field is related to the dimension $\D$ of the dual
operator by $m^2 = (\D-d) \D$.
The bulk field equations are given by
\be \label{bulk}
G_{\m \n} = \k^2 \tilde{T}_{\m \n}(\F), \qquad \Box_g \F = \pa V/\pa \F
\ee
where $G_{\m \n}$ is the Einstein tensor, 
$\Box_g \F= \frac{1}{\sqrt{g}} \pa_\m (\sqrt{g} g^{\m \n} \pa_\n \F)$
 and $\tilde{T}_{\m \n}(\F)$ is the
stress energy tensor associated with the scalar field $\F$ (see (\ref{defT})).
 
The method of holographic renormalization now consists of the
following steps (a more detailed discussion can be found 
in \cite{Skenderis:2002wp}).

{\bf 1. Asymptotic solutions}

In the first step one works out the most general asymptotic solutions
with given Dirichlet data
\bea
ds^2 &=& \frac{d \r^2}{ 4 \r^2} + \frac{1}{\r} g_{ij}(x,\r) dx^i dx^j  \\
\F(x,\r) &=& \r^{(d-\D)/2} \f(x,\r)
\eea
where\footnote{In general, the expansions may 
involve half integral powers of $\r$. In such cases it is more
natural to use a new radial coordinate $r$, where $\r=r^2$.},
\bea
g_{ij}(x,\r) &=& g_{(0)ij} + \r g_{(2)ij} + \cdots + \r^{d/2}(g_{(d)ij}
+ \log \r  h_{(d)ij} ) + \cdots \\
\f(x,\r) &=& \f_{(0)} + \r \f_{(2)} + \cdots + \r^{\D - d/2}
(\f_{(2 \D -d)} + \log \r  \psi_{(2 \D -d)}) + \cdots  
\eea
In this expansion, $g_{(0)ij}$ and $\f_{(0)}$ are identified with the
QFT sources that couple to the dual operators, as discussed in the
previous section.

Inserting these expansions in the bulk field equations (\ref{bulk})
one finds that all coefficients but $\f_{(2 \D -d)}$ and the 
traceless transverse part of $g_{(d)ij}$
are locally determined by $g_{(0)ij}$ and $\f_{(0)}$ 
\cite{FeffermanGraham,HS,dHSS} (see the appendices of \cite{dHSS}
for explicit expressions of the coefficients).
The part of  $g_{(d)ij}$ that is determined, i.e.
$\nabla^i g_{(d)ij}$ and $\Tr g_{(d)}$,
encodes Ward identities and anomalies, as we discuss below.
We will call  $g_{(d)ij}$ and $\f_{(2 \D -d)}$ the response
functions. The logarithmic
terms appear only in special cases: $h_{(d)}$ only in even dimensions
and $\psi_{(2 \D-d)}$ only when $\D - d/2$ is an integer. Both of
them are directly related to the conformal anomalies discussed
in the previous section: $h_{(d)}$ is the metric variation 
of the gravitational part of the conformal anomaly and
$\psi_{(2 \D-d)}$ is the variation w.r.t. $\f_{(0)}$ of the
matter part of the conformal anomaly \cite{dHSS}. 

{\bf 2. On-shell divergences}

Having obtained the asymptotic solutions we now obtain the most
general divergences of the on-shell action,
\be
S_{reg}[g_{(0)},\f_{(0)};\e]=\int d^dx \sqrt{g_{(0)}}
\left(\sum_{\nu} a_{(\n)} \e^{-\n} - \log \e a_{(d)} \right) + \co(\e)
\ee
where $\e$ is a cut-off in the radial coordinate, $\r \geq \e$.
It turns out all coefficients $a_{(\n)}$ depend only
on $g_{(0)}$ and $\f_{(0)}$ but not on the undetermined coefficients
$g_{(d)}$ and $\f_{(2 \D -d)}$. The coefficient $a_{(d)}$ is equal
to the conformal anomaly of the dual CFT \cite{HS}.

{\bf 3. Counterterms and renormalized action}

To obtain a well-defined on-shell action we should
subtract the infinities and then remove the regulator.
To do this we first express the divergent terms found
in the previous step in terms of induced fields at
the hypersurface $\r=\e$. This entails inverting
the asymptotic series obtained in the first step
and inserting it in the divergent terms obtained
in the second step. This is one of the most laborious steps of the 
procedure. The end result is the counterterm action, $S_{ct}$. 
The renormalized action
is defined by
\be
S_{ren}=\lim_{\e \to 0} S_{sub}, \qquad S_{sub}= S_{reg}+S_{ct}.
\ee

{\bf 4. 1-point functions in the presence of source}

We can now differentiate the renormalized action to
obtain the 1-point function in the presence of
sources \cite{dHSS}, 
\bea \label{1pt}
\langle T_{ij}(x) \rangle_s \equiv \frac{2}{ \sqrt{g_{(0)}(x)}} 
\frac{ \delta S_{ren}}{\delta g^{ij}_{(0)}(x)} &=&
\lim_{\e \to 0} \left(\frac{1}{\e^{d/2-1}}\frac{2}{\sqrt{\g(x,\e)}} 
\frac{ \delta S_{sub}}{\delta \g^{ij}(x,\e)}  \right) \nonumber \\
&=& \frac{d}{2 \k^2} g_{(d)ij}
+ X[g_{(0)},\f_{(0)}], \\
\langle O(x) \rangle_s \equiv \frac{1}{\sqrt{g_{(0)}(x)}} 
\frac{ \delta S_{ren}}{\delta \f_{(0)}(x)} &=&\lim_{\e \to 0}
\left(\frac{1}{\e^{\D/2}}\frac{1}{\sqrt{\g(x,\e)}} 
\frac{ \delta S_{sub}}{\delta \F(x,\e)}\right) \nonumber \\
&=& (d-2 \D) \phi_{(2 \D -d)}
+ Y[g_{(0)},\f_{(0)}] \nonumber
\eea
where $X[g_{(0)},\f_{(0)}]$ and $Y[g_{(0)},\f_{(0)}]$ are (known)
local expressions that depend on sources. The first equality 
is a definition. In the second equality we expressed the 
1-point function as a limit of the regulated 1-point function.
The regulated 1-point function can be computed in all 
generality and the limit can be explicitly taken. This 
is a straightforward but rather tedious computation.
The result is the one shown above. We thus find that the correlation
functions depend on the coefficient that the asymptotic analysis
left undetermined. As discussed above, the near boundary analysis
does determine the divergence and trace of $g_{(d)ij}$. This means that
the divergence and trace of $\langle T_{ij}(x) \rangle_s$ can be
determined. This yields the Ward identities, including anomalies,
that we discussed in the previous section. 
The relations (\ref{1pt}) imply that the pairs $(g_{(0)}, g_{(d)})$
and $(\f_{(0)}, \f_{(2 \D - d)})$ are conjugate pairs.

{\bf 5. Correlation functions}

To obtain higher point functions we should further differentiate
(\ref{1pt}) w.r.t. the sources. The expressions
$X[g_{(0)},\f_{(0)}]$ and $Y[g_{(0)},\f_{(0)}]$
lead to only local contributions. The (non-local) $n$-point function
is thus encoded in the dependence of $ g_{(d)}$ and $\f_{(2 \D - d)}$
on the sources. We thus reach the conclusion:
\\

\framebox[\width]{
\begin{minipage}{6in}
\hspace{.2in} The theory is solved if we determine the response
functions in terms of the sources.
\end{minipage} 
}
\\

To obtain such a relation we need  a regular exact
(as opposed to asymptotic) solution of the bulk equations
with the boundary conditions specified by the sources.
In the absence of more powerful methods one can proceed
perturbatively. One can determine the
response functions to linear order by solving
the bulk field equations linearized around a background 
solution \cite{BFS}. (The background solution specifies the 
vacuum of the dual QFT, see section 6.1 of \cite{Skenderis:2002wp}).
Higher-point functions can be computed  by solving the
bulk equations perturbatively in a bulk coupling
constant. Examples have been discussed in \cite{Skenderis:2002wp,3pt}.

The procedure described here is general and can be carried
out in all cases. The steps however appear to have
certain redundancy. In step 1 and 2 the asymptotic
solution and divergences are obtained in terms of the
Dirichlet data. In order to obtain the counterterms
however
one should invert the asymptotic series. Then the
1-point functions are obtained in terms of the
induced fields at $\r=\e$ and the asymptotic solution is
used again to obtain the final expression for
1-point functions. Clearly, it would be desirable
to avoid having to go back and forth from asymptotic
data to covariant fields. A related issue is the following.
In step 2 we mentioned that the divergences depend only
on the sources but not the response functions. This
followed from an explicit computation. It would be more
satisfying to make this manifest. We discuss in the
next section an approach that removes these drawbacks.
A related work that also leads to simplifications 
can be found in \cite{Martelli:2002sp}.

\section{Hamiltonian approach to Holographic Renormalization}

Let $\overline{M}$ be a conformally compact, Riemannian (d+1)-manifold, 
$M$ its 
interior and $\partial M$ its boundary. We will consider the following 
action for the 
Riemannian metric $g_{\m\n}$ on $\overline{M}$ 
\beq \label{act} 
S_{gr}[g]=-\frac{1}{2\kappa^2}\left[\int_M d^{d+1}x\sqrt{g}R+
	\int_{\partial M}d^dx\sqrt{\g}2K\right],
\eeq 
where $\kappa^2=8\pi G_{d+1}$, $\g$ is the induced metric
on $\partial M$ and  $K$ is the trace of the extrinsic curvature of
 the boundary. This is the standard Einstein-Hilbert action with the Gibbons-Hawking 
boundary term which ensures that the variational problem is well-defined.
The overall sign is chosen 
so that the action is positive definite when evaluated on a classical solution
in the vicinity of (Euclidean) AdS. 
\footnote{
Our convention for the Riemann tensor is
$R^\m\phantom{}_{\r\n\s}=\partial_\n \G^\m_{\r\s}+\G^\m_{\l\n}\G^\l_{\r\s}-
(\n\leftrightarrow\s)$. This differs by an overall sign from the conventions
used in \cite{HS,dHSS}.} To allow for  matter we add  
\beq 
S_{m}=\int_M d^{d+1}x\sqrt{g}\mathcal{L}_m.
\eeq
to the gravitational action, where $\mathcal{L}_m$ is a generic matter 
field Lagrangian density. The stress tensor is 
then defined in the standard fashion by 
\beq \label{defT}
\d_g S_m\equiv\frac 12 \int_M d^{d+1}x\sqrt{g} \tilde{T}_{\mu\nu}\d g^{\m\n}.
\eeq
The Euler-Lagrange equations of the total action $S=S_{gr}+S_m$ are  Einstein's equations 
\beq \label{Einstein}
 G_{\mu\nu}=\kappa^2 \tilde{T}_{\mu\nu}
\eeq 
and the matter field equations.

Our method of holographic renormalization makes use of the ADM formalism and the Gauss-
Codacci equations which we will now briefly review. The standard ADM 
formalism (see, for instance, \cite{Wald}) for a pseudo-Riemannian manifold relies 
on the existence of a
{\em global} time function $t$ which is used to foliate space-time into diffeomorphic
hypersurfaces of constant $t$. For a generic Riemannian manifold, however, there is no 
natural choice of time as all coordinates are equivalent. Nevertheless, for a Riemannian 
manifold with boundary one can use the coordinate `normal' to the boundary as a global
`time' coordinate and, hence, foliate the manifold into hypersurfaces diffeomorphic to
the boundary. For asymptotically (Euclidean) AdS manifolds this can always be done at 
least in a neighborhood of the boundary \cite{FeffermanGraham} (see also
the recent review \cite{anderson} and references therein). The question of if and 
where this `radial' coordinate emanating from the boundary ceases to be well-defined 
depends on the topology of the space and will not be addressed here.

Let $r$ be the `radial' coordinate emanating from the boundary of a Riemannian 
manifold with boundary $(M,g_{\m\n})$ in the way described above and consider the 
hypersurfaces $\S_r$ defined by $r(x)={\rm constant}$. The unit normal to $\S_r$,
pointing in the direction of increasing $r$, is given by $n^\m=\frac{1}{\parallel 
dr\parallel_g} g^{\m\n}\frac{\partial r}{\partial x^\n}\mid_\S$. This allows one 
to express the induced metric on the hypersurfaces in a coordinate independent
fashion as\footnote{We use a hat to denote tensors that are purely transverse to the
unit normal, i.e. quantities which vanish when contracted with $n^\m$.} 
 $\hat{\g}_{\m\n}= g_{\m\n}-n_\m n_\n $. The metric on $M$ can then be decomposed 
as
\beq 
ds^2=g_{\m\n}dx^\m dx^\n=\hat{\g}_{\m\n}d\hat{x}^\m d\hat{x}^\n+2N_\m
	 d\hat{x}^\m d r+(N^2+N_\m N^\m)d r^2
\eeq
where $N$ and $N^\m$ are respectively the {\em lapse function} and the 
{\em shift function}. They correspond to non-dynamical degrees of freedom which
we will `gauge-fix' shortly. Geometrically they measure how `normal'  the 
coordinate $r$ is to the hypersurfaces: the choice $N=1$, $N^\m=0$ makes $r$ a
Gaussian normal coordinate, in which case $n^\m$ becomes tangent to {\em geodesics}
normal to the hypersurfaces. A quantity that will be of central importance in our 
analysis is the extrinsic curvature of the hypersurfaces
\beq 
\hat{K}_{\m\n}=\hat{\g}_\m^\r\nabla_\r n_\n=\frac 12 \pounds_n \hat{\g}_{\m\n},
\eeq
where $\pounds_n$ denotes the Lie derivative with respect to the unit 
normal $n^\m$. Thus, the extrinsic curvature measures the radial evolution
of the induced metric and hence encapsulates all  dynamical information of
the geometry of the hypersurfaces. In fact, the Riemann tensor of the d+1 
dimensional manifold $M$ can be expressed entirely in terms of the intrinsic 
(i.e. Riemannian) and extrinsic curvatures of the hypersurfaces $\S_r$ via the
so called {\em Gauss-Codacci} equations
\beqa \label{GC} 
\hat{\g}^\a_\m\hat{\g}^\b_\n\hat{\g}^\g_\r\hat{\g}^\d_\s R_{\a\b\g\d}=
	\hat{R}_{\m\n\r\s} + \hat{K}_{\m\s}\hat{K}_{\n\r}-\hat{K}_{\m\r}
	\hat{K}_{\n\s}\\
\hat{\g}^\r_\n n^\s R_{\r\s}=\hat{\nabla}_\m \hat{K}^\m_\n-\hat{\nabla}_\n
	 \hat{K}^\m_\m. \nonumber
\eeqa 
These purely geometric equations exhibit most explicitly the implications of
the bulk (d+1)-dimensional geometry for the geometry of the hypersurfaces. For,
instance, one sees immediately that conformal flatness of the bulk manifold
implies very strong constraints on the extrinsic curvature of the radial slices.
The case of interest to us here is, of course, the case of an Einstein bulk 
manifold. A little manipulation of the Gauss-Codacci equations brings them
in the following form, most suitable to exhibit the consequences of $M$ being 
Einstein - we stress that the following equations are purely geometric: 
\beqa
\label{einsteinI}\hat{K}^2-\hat{K}_{\m\n}\hat{K}^{\m\n}=\hat{R}+2G_{\m\n}n^\m n^\n, 
\nonumber \\
\label{einsteinII}\hat{\nabla}_\m \hat{K}^\m_\n-\hat{\nabla}_\n \hat{K}^\m_\m=G_{\r\s}
	\hat{\g}^\r_\n n^\s, \\
\label{einsteinIII}\pounds_n \hat{K}_{\m\n}+\hat{K}\hat{K}_{\m\n}-2\hat{K}_{\m}
	\phantom{}^\r\hat{K}_{\r\n}=\hat{R}_{\m\n}-\hat{\g}^\r_\m\hat{\g}^\s_\n 
	R_{\r\s}. \nonumber
\eeqa
These equations 
become dynamical once we use Einstein's equations to replace the Einstein tensor
with the matter stress tensor.
When $M$ is both conformally flat and Einstein they can be solved 
exactly \cite{Skenderis:1999nb}. Note that conformal flatness is automatic 
if $M$ is three dimensional. 
However, solving these equations in general for an arbitrary Einstein manifold is 
equivalent to solving Einstein's equations and, therefore, far from trivial.

The ADM formalism allows us to express the bulk action in terms of transverse
quantities as
\beq 
S=-\frac{1}{2\k^2}\int_Md^{d+1}x\sqrt{\hat{\g}}N(\hat{R}+\hat{K}^2-\hat{K}
	_{\m\n}\hat{K}^{\m\n}-2\k^2\mathcal{L}_m).
\eeq
The canonical momenta can now be defined in the standard fashion\footnote{
One may consider adding extra {\it finite} local 
boundary terms in the action (\ref{act}).
These would result in additional terms in the momenta and finally lead
to additional contact terms in (holographically computed) correlators. 
The addition of such boundary 
terms is the counterpart of finite local counterterms related to
the scheme dependence of the boundary QFT.}
\beq\label{momentaI} 
\p^{\m\n}\equiv \frac{\d L}{\d \dot{\hat{\g}}_{\m\n}}=
	-\frac{1}{2\k^2}\sqrt{\hat{\g}}\left(\hat{K}\hat{\g}^{\m\n}-
	\hat{K}^{\m\n}\right),\qquad \p^I\equiv \frac{\d L}{\d \dot{\F}_I}\,,
\eeq 
where $\F_I$ is a generic matter field and the Lagrangian $L$ is defined as usual by
$S=\int dr L$. In particular, the canonical momenta
conjugate to the lapse and shift functions vanish identically, and hence the
corresponding equations of motion in the canonical formalism become constraints, 
which are precisely the first two equations in (\ref{einsteinII}). 

Let us finally consider the on-shell gravitational action, as it is precisely 
this quantity that becomes the generating functional of connected correlation 
functions of the dual field theory on the boundary. From Einstein's equation
and (\ref{einsteinI}) it follows that
\beq 
S_{\rm on-shell}=-\frac{1}{\k^2}\int_{r=r_0}^{r=r_1}drd^{d}x\sqrt{\hat{\g}}N\left[\hat{R}+
	\k^2\left(n^\m n^\n \tilde{T}_{\m\n}-\mathcal{L}_m\right)\right],
\eeq
where the boundary is located at $r=r_1$ and $r_0$ ($r_0<r_1$ by our definition 
of the unit normal) defines a hypersurface in the 
interior of $M$. As mentioned above, there always exists an $r_0$ sufficiently close
to $r_1$ such that the above expression for the on-shell action is well-defined, but
there may not exist an $r_0$ such that the integration from $r_0$ to $r_1$ covers the
entire manifold. However, this issue is irrelevant for 
the near boundary analysis. The on-shell action is a functional of the boundary
values of the fields $\hat{\g}(r_1,x)$ and $\F_I(r_1,x)$. The corresponding momenta
on $\S_{r_1}$ are then obtained from the on-shell action by
\beq\label{momenta} 
\p^{\m\n}(r_1,x)= \frac{\d S_{\rm on-shell}}{\d\hat{\g}_{\m\n}(r_1,x)},\qquad
	\p^I(r_1,x)=\frac{\d S_{\rm on-shell}}{\d \F_I(r_1,x)}.
\eeq   
However, asymptotically AdS spaces are non-compact and the boundary is located at 
$r_1=\infty$. So the above expressions for the on-shell action and canonical 
momenta evaluated
at $r_1$ contain divergences due to the infinite volume of the bulk manifold. 
The advantage of the Hamiltonian formulation is that these expressions hold 
identically for any hypersurface $\S_r$ defined by any {\em finite} $r$. 

This is then a good point to detail the philosophy of the new approach.
\begin{enumerate}
\item Using the Hamiltonian formalism we have arrived at a manifestly
	covariant expression for the canonical momenta evaluated on an 
	arbitrary hypersurface $\S_r$ - for finite $r$. In particular,
	the momenta are {\em functionals of the bulk fields on} $\S_r$.
\item Using Einstein's equations in the Gauss-Codacci relations 
	(\ref{einsteinII}) -  
	together with any
	extra equations of motion for matter fields - we obtain a set of
	second  order differential equations for the induced metric and 
	the other bulk fields evaluated on $\S_r$.
\item This set of second order ordinary differential equations is then 
	turned into a set of {\em first order functional partial differential 
	equations} by expressing the radial derivative as a functional 
	derivative. The crucial point here is that the canonical 
	momenta are essentially the $r$-derivative of the corresponding
	bulk fields (up to issues relating to gauge fixing to be discussed
	 below) and we have just seen that the momenta  are functionals of the
	bulk fields on $\S_r$. Hence,
	\beq
	\frac{d}{dr}=\int d^dx
		2\hat{K}_{\m\n}[\hat{\g},\F]\frac{\d}{\d\hat{\g}_{\m\n}}+
		\int d^dx
		\dot{\F}_I[\hat{\g},\F]\frac{\d}{\d\F_I}
	\eeq
	 This step is reminiscent, essentially equivalent, to the theorem of 
	Jacobi \cite{CourantHilbert} in the Hamilton-Jacobi theory of classical
	mechanics, where one expresses the momenta as functional derivatives
	of the on-shell action as above, but then derives a partial 
	differential equation for the on-shell action. This is
	precisely the approach followed in \cite{deBoer:1999xf}. However, we
	derive functional PDE's for the momenta and this is 
	advantageous as we will discuss momentarily.
\item The set of first order functional PDE's thus obtained are, of course,
	much harder to solve than the original set of second order 
	ODE's\footnote{In classical mechanics - where the PDE's are not
	functional -  it is often easier to solve the PDE's (either for 
	the action, i.e. the Hamilton-Jacobi equation, or for the momenta)
	and, as a result solve Hamilton's equations. This amounts to the
	`inverse method of characteristics', as Hamilton's equations are
	just the characteristic equations for the Hamilton-Jacobi equation.},
	but this representation of the problem is most suitable for the 
	near-boundary analysis in asymptotically AdS spaces, where the bulk
	fields satisfy prescribed but arbitrary Dirichlet boundary conditions:
	\beq
	 \hat{\g}_{\m\n}\sim {\rm e}^{2r}\hat{\g}\sub{0}_{\m\n}(x),\,\,\,
		\F_I\sim {\rm e}^{(\D_I-d)r}\f\sub{0}_I(x)
	\eeq
	as $r\to\infty$, where $\D_I$ is the scaling dimension of the operator
	dual to the bulk field $\F_I$. Provided these asymptotics hold\footnote{
	When $\D=d/2$ the leading asymptotics of the bulk fields
	are of the form $r \exp (-dr/2)$. In those  cases the 
	functional representation of the radial derivative must be modified 
	\cite{IP&KS}, but the above procedure for the asymptotic analysis applies 
	equally well.}, the asymptotic form of the functional representation 
	of the radial derivative is very suggestive:
	\beq
	\partial_r\sim \int d^dx
		2\hat{\g}_{\m\n}\frac{\d}{\d\hat{\g}_{\m\n}}+
		\int d^dx
		(\D_I-d)\F_I\frac{\d}{\d\F_I}.
	\eeq
	Not surprisingly, this is the total dilatation operator, $\d_D$, of the 
	theory, 
	which appears as a consequence of the well-defined scale transformation
	rules the fields obey asymptotically. 
	From the point of view of the 
	boundary field theory, this is precisely the Callan-Symanzik equation
	obeyed by the renormalized one-point functions in the presence of sources 
	- which, as we will see, 
	are related to the canonical momenta. In the spirit of perturbation 
	theory then, it is natural to {\em expand the momenta in eigenfunctions
	of the total dilatation operator} and solve the functional PDE's
	`perturbatively', i.e. asymptotically, while preserving covariance. 
	This is in contrast to the method of holographic renormalization, where
	one is seeking asymptotic expansions of the bulk fields using the distance
	from the AdS boundary as a small parameter - thus
	explicitly breaking bulk covariance.  
\item 
	In contrast to previous methods, our focus here is on the canonical 
	momenta and {\em not} the on-shell action. From the field theory point 
	of view, this is to say we are interested in the exact one-point
	functions - as opposed to the partition function in the presence of
	sources. The connection between the canonical momenta and the one-point
	functions is surprisingly simple in our formalism. On the regulating 
	surface $\S_r$ with $r$ finite, the (unrenormalized but regulated) 
	one-point functions
	are given by the AdS/CFT prescription as the functional derivative of the 
	on-shell action with respect to the corresponding source, i.e. bulk field.
        Let us show this for the case of matter fields (the derivation
	for gravity is the same). The variation of the action
	is given by,
\be
\d S = \frac{\pa L}{\pa \dot{\Phi}_I} \delta \Phi_I |^{r}
+\int^{r} dr \delta \Phi_I \left[\frac{\pa L}{\pa \F_I} 
- \pa_r \left(\frac{\pa L}{\pa \dot{\F}_I}\right)\right], 
\ee     
        where we assumed (without loss of generality) 
	that the matter fields have a standard kinetic term and
	we gauge fixed as in (\ref{gf_metric}). The second term is just the Euler-Lagrange
	equation and thus vanishes on-shell. 
	We therefore obtain,
\be
\frac{\d S_{{\rm on-shell}}}{\d \F_I} = \p^I(r),
\ee
        where we used (\ref{momentaI}). The l.h.s. is, by the AdS/CFT dictionary, the 
	regulated 1-point function in the presence of sources.
	We now argue that a similar 
	connection holds for the {\em renormalized} one-point functions as well.
	The renormalized one-point functions are defined to be the one-point 
	functions one obtains from the renormalized action. This is in turn the 
	on-shell action plus a set of covariant counterterms which remove the 
	divergences of the on-shell action as $r\to\infty$. Suppose now these
	covariant counterterms for the on-shell action are constructed. Taking 
	the functional derivative with respect to the appropriate bulk field
	they lead to covariant terms which when added to the regularized momenta
	must - by construction - remove all the potential singularities from the
	canonical momenta. We thus need to identify the singular
	part of canonical momentum and remove it. The main constraint is that 
	the subtraction should be covariant.
	This is done by expanding the momentum  in terms of (covariant) eigenfunctions 
	of the dilatation operator and observing that the divergent terms 
	have eigenvalues less than the dimension $\D$ of the dual operator.
	To summarize, we have the very general result
	\beq \label{ren1pt}
	\framebox[\width]{
	\begin{minipage}{5.5in}
	\begin{eqnarray*} \langle \hat{T}_{\m\n}\rangle_{\rm ren}=-\frac{1}{\k^2}\left(
		\hat{K}\sub{d}_{\m\n}-\hat{K}\sub{d}\hat{\g}_{\m\n}\right),
		\phantom{more space}
		\langle \mathcal{O}_I\rangle_{\rm ren}=\frac{1}{\sqrt{\hat{\g}}}
		\p^I\sub{\D_I}, \\ 
	\end{eqnarray*}
	\end{minipage}}
	\eeq
	The terms on the right hand sides have the (engineering) dimension indicated by 
	their subscript. These would have also been their dilatation eigenvalues   
	in the absence of conformal anomalies and RG running. In most 
	examples, the bulk theory involves non-trivially only fields of the gauged 
        supergravity obtained by reducing the 10d supergravity over a compact manifold.
	In such cases, the bulk fields correspond to operators of protected dimensions
	and thus the coefficients in (\ref{ren1pt}) fail to be eigenfunctions of the 
	dilatation operator only because of the conformal anomaly. As we shall see 
	explicitly in the examples, the conformal anomaly induces an inhomogeneous 
	term in the dilatation transformation of the coefficients in (\ref{ren1pt}). 
	We further note that these coefficients are not completely determined by the 
	asymptotic analysis
        -- they are the counterparts of the undetermined coefficients of the 
	near-boundary analysis. It is therefore redundant to first construct
	covariant counterterms for the on-shell action and then use them to obtain
	the renormalized one-point functions, since the equations of motion can be 
	solved for the momenta directly - and these are all one needs to obtain
	the one-point functions. 
\item   Although, as we just argued, it is not necessary to compute covariant counterterms
        for the on-shell action in order to obtain renormalized correlation functions,
	one {\it can} construct them as a secondary step in our
	method, and in fact more efficiently than previous methods. This is done
	by constructing a differential equation - essentially equivalent to 
	the Hamilton-Jacobi equation - for the on-shell action which we then solve
	in parallel to the equations for the momenta. Explicit examples will 
	be presented below. 

	In previous works where a Hamiltonian approach was used 
	\cite{deBoer:1999xf}-\cite{Fukuma:2002sb}  
	a central point of the analysis was the solution of the Hamilton-Jacobi (HJ) 
	equation for the on-shell action. In this context, the HJ equation
	is a functional PDE for 
	the on-shell action which can be solved by inserting an ansatz for the 
	on-shell action in it. By requiring that terms with different number of 
	derivatives cancel separately one gets a number of equations,
	the descent equations, that can be solved to determine the unknown functions
	in the ansatz. In the presence of scalars, the equations were further 
	organized in \cite{Martelli:2002sp} according to the number of scalar fields they contain. 
	The resulting equations are not in general equivalent to the ones 
	in the standard approach. This is due to the fact that the scalar 
	fields are treated differently than in the standard holographic renormalization. 
	Recall that in the standard approach the equations are solved by using the 
	distance from the boundary as a small parameter with all sources being unconstrained.
	The expansion in the number of scalar
	fields requires that all scalar fields are (equally) small and for 
	this to be the case the Dirichlet data (QFT sources) should be tuned
	to be (appropriately) small. This is rather unnatural 
	since on the QFT side all sources are unconstrained and of order one and in general
        can lead to erroneous results. In simple examples,
	such as the ones studied in the literature, there is no obstruction
	in considering the sources small and the results so obtained are in agreement
	with results obtained via the standard method. An alternative approach 
	that overcomes these issues is to organize the terms in the HJ 
	equation according to their dilatation weight. This yields equations
	that are equivalent to the ones in the standard holographic renormalization
	method. Solving the HJ equation for the on-shell action leads to  
	some of the same simplifications we find here. For instance,
	the covariant counterterm action is derived easier. On the other hand,
	the use of an ansatz for the on-shell action 
	(instead of constructively obtaining the most general solution)
	as well as various sign ambiguities make the method less rigorous than
	the standard approach. More importantly, 
	focusing on the canonical momenta instead of the on-shell action
	appears to be the most economic way to proceed.
\item Apart from an elegant framework for the general asymptotic analysis, this
	formalism provides a most efficient way to calculate {\em correlation functions}
	of the boundary field theory holographically. As we have just seen 
	this amounts to determining the renormalized canonical momenta as functionals of
	arbitrary bulk fields - i.e. as functionals of arbitrary sources. To determine
	2-point functions we only need to determine the momenta in terms of the source
        at linearized level. Furthermore, the contribution
	of the counterterms to 2-point functions can also be determined directly  
	from the {\it linearized} analysis, following the discussion in the previous 
	point. A similar discussion applies also to $n$-point functions ($n>2$).
	This leads to a significant simplification of the computation of 
	correlation functions. Details will appear elsewhere \cite{IP&KS}.

\end{enumerate}
	   
{\em Gauge fixing}

Before we carry out the near boundary analysis for pure AdS gravity and gravity
coupled to scalars following the above prescription, let us fix the gauge freedom
associated with the shift and lapse functions by setting $N^\m=0$ and $N=1$. The 
bulk metric then takes the form\footnote{All tensors are transverse and so we 
drop the hats form now on.}  
\beq \label{gf_metric}
ds^2=d r^2+\g_{ij}( r,x)dx^idx^j,
\eeq 
where $i,\,j=1,\ldots d$ are indices along the hypersurfaces. The extrinsic
curvature becomes 
\beq 
K_{ij}=\frac12 \dot{\g}_{ij}
\eeq 
where the dot denotes differentiation with respect to $r$. The non-vanishing
components of the Christoffel symbol are
\beq 
\G^{d+1}_{ij}=-K_{ij},\;\;\;\;\;\G^i_{d+1j}= K^i_j,
	\;\;\;\;\;\G^i_{jk}.
\eeq 
The gravitational field equations (\ref{einsteinII}) take the form
\beqa 
K^2-K_{ij}K^{ij}=R+2\k^2 \tilde{T}_{d+1d+1},\nonumber \\
\label{einsteinIIgf}\nabla_iK^i_j-\nabla_jK=\k^2 \tilde{T}_{jd+1},\\
\dot{K}^i_j+KK^i_j=R^i_j-\k^2\left(\tilde{T}^i_j+\frac{1}
	{1-d}\tilde{T}^\s_\s\d^i_j\right).\nonumber
\eeqa 
$\dot{K}^i_j$ here stands for $\frac{d}{dr}(\g^{ik}K_{kj})$. An additional 
equation for the on-shell action can be derived as promised. Since, 
\beq 
\dot{S}_{\rm on-shell}=L=-\frac{1}{\k^2}\int_{\S_ r}d^dx\sqrt{\g}\left[R+\k^2(
	\tilde{T}_{d+1d+1}-\mathcal{L}_m)\right],
\eeq    
we can obtain an expression for $S_{\rm on-shell}$ if we write the integrand as the
derivative of some (covariant) quantity. This is achieved by introducing a 
covariant variable $\l$ and writing
\beq \label{on_shell_action}
S_{\rm on-shell}=-\frac{1}{\k^2}\int_{\S_{r}}d^dx\sqrt{\g}(K-\l).
\eeq 
Taking the trace of the third equation in (\ref{einsteinIIgf}) we determine that $\l$
satisfies 
\beq \label{lambda} 
\dot{\l}+K\l-\k^2\left(\mathcal{L}_m-\frac{1}{1-d}\tilde{T}^\s_\s\right)=0.
\eeq 

\subsection{Pure gravity case}

We will now demonstrate the method of Hamiltonian holographic renormalization
for pure gravity with a negative cosmological constant\footnote{$S\sim\int (R-2\L)$.}
 $\L=d(1-d)/2$. The equations of motion reduce to 
\beqa 
K^2-K_{ij}K^{ij}=R+d(d-1),\nonumber \\
\label{einsteinIIgr}\nabla_iK^i_j-\nabla_jK=0,\\
\dot{K}^i_j+KK^i_j=R^i_j+d\d^i_j. \nonumber
\eeqa
The on-shell action is determined from the equation
\beq \label{lambda_gr} 
\dot{\l}+K\l=d.
\eeq 

We will expand the extrinsic curvature and $\l$ in eigenfunctions of the 
dilatation operator, which now takes the form
\beq
\d_D=\int d^dx2\g_{ij}\frac{\d}{\d\g_{ij}}.
\eeq
Then,
\beqa\label{cov_expansions} 
K^i_j[\g]=K\sub{0}^i_j+K\sub{2}^i_j+\cdots+K\sub{d}^i_j+\tilde{K}
	\sub{d}^i_j\log{\rm e}^{-2r}+\cdots, \nonumber \\  
\l[\g]=\l\sub{0}+\l\sub{2}+\cdots+\l\sub{d}+\tilde{\l}\sub{d}\log{\rm e}^{-2r}+\cdots,
\eeqa 
where
\beqa 
\d_DK\sub{n}^i_j=-n K\sub{n}^i_j,\,\,n<d,\,\,\,\d_D\tilde{K}\sub{d}^i_j=
	-d \tilde{K}\sub{d}^i_j,\nonumber \\
\d_DK\sub{d}^i_j=-d K\sub{d}^i_j-2\tilde{K}\sub{d}^i_j \\
\d_D\l\sub{n}=-n \l\sub{n},\,\,n<d,\,\,\,\d_D\tilde{\l}\sub{d}=
	-d \tilde{\l}\sub{d},\nonumber \\
\d_D\l\sub{d}=-d\l\sub{d}-2\tilde{\l}\sub{d}. \nonumber
\eeqa
The inhomogeneous transformations of $K\sub{d}^i_j$ and $\l\sub{d}$, which follow 
immediately from the relation between the radial derivative and the dilatation operator,
are due to the conformal anomaly.

Before we proceed to determine these coefficients from the equations of motion,
let us exhibit the equivalence of this covariant expansion in eigenfunctions
of the dilatation operator to the asymptotic expansion of the induced metric in 
the standard holographic renormalization method. There the induced metric is
expanded in $\r=\exp(-2r)$ as
\beq\label{metric_expansion}
\g_{ij}=\frac{1}{\r}\left[g\sub{0}_{ij}+\r g\sub{2}_{ij}+\cdots+\r^{d/2}
	g\sub{d}_{ij}+\r^{d/2}\log\r h\sub{d}_{ij}+\cdots\right]
\eeq
Hence
\beq\label{dotmetric_expansion} 
\frac12\dot{\g}_{ij}=\frac1\r g\sub{0}_{ij}-\r g\sub{4}_{ij}+\cdots
	+\r^{(d/2-1)}\left[\left(1-\frac d2\right)g\sub{d}_{ij}-
	h\sub{d}_{ij}\right]+\r^{(d/2-1)}\log\r\left(1-\frac d2\right)
	h\sub{d}_{ij}+\cdots
\eeq
However, each term in the covariant expansion of the extrinsic curvature is
a functional of the induced metric $\g_{ij}$. Using the expansion 
(\ref{metric_expansion}) of the metric we can functionally expand the
eigenfunctions of the dilatation operator as
\beqa
K\sub{0}_{ij}[\g]=\g_{ij}=\frac{1}{\r}\left[g\sub{0}_{ij}+\r g\sub{2}_{ij}+
	\cdots+\r^{d/2}g\sub{d}_{ij}+\r^{d/2}\log\r h\sub{d}_{ij}+
	\cdots\right] \nonumber \\
K\sub{2}_{ij}[\g]=K\sub{2}_{ij}[g\sub{0}]+\r
	\int d^dx g\sub{2}_{kl}\frac{\d K\sub{2}_{ij}}
	{\d g\sub{0}_{kl}}+\cdots  \nonumber \\ 
\vdots\phantom{more space here}  \\
K\sub{d}_{ij}[\g]=\r^{(d/2-1)}K\sub{d}_{ij}[g\sub{0}]+\cdots \nonumber \\
\tilde{K}\sub{d}_{ij}[\g]=\r^{(d/2-1)}\tilde{K}\sub{d}_{ij}[g\sub{0}]+\cdots \nonumber
\eeqa
Inserting these expressions in the covariant expansion for $K_{ij}$ and
comparing with (\ref{dotmetric_expansion}) we determine
\beqa \label{relations} 
K\sub{0}_{ij}[g\sub{0}]=g\sub{0}_{ij},\phantom{more space here}\\ \NO 
	K\sub{2}_{ij}[g\sub{0}]=-g\sub{2}_{ij}[g\sub{0}],
	\phantom{more space here}\\ \NO \vdots \phantom{even more more
	 space here}\\ \NO K\sub{n}_{ij}[g\sub{0}]=-\frac n2 g\sub{n}_{ij}
	[g\sub{0}]+{\rm lower},\phantom{more space here}\\ \NO
	 \vdots\phantom{even more more space here} \\ \NO K\sub{d}_{ij}
	[g\sub{0}]=-\frac d 2 g\sub{d}_{ij}[g\sub{0}]-h\sub{d}_{ij}
	[g\sub{0}]+{\rm lower},\phantom{more space here}\\ \NO \tilde{K}
	\sub{d}_{ij}[g\sub{0}]=-\frac d 2 h\sub{d}_{ij}[g\sub{0}],
	\phantom{more space here}
\eeqa 
where `lower' stands for terms involving functional derivatives with respect 
to $g\sub{0}_{ij}$ of lower order coefficients $g\sub{k}_{ij}[g\sub{0}]$.
For d=4, for example, 
\beq 
K\sub{4}_{ij}[g\sub{0}]=-2 g\sub{4}_{ij}[g\sub{0}]-h\sub{4}_{ij}
	[g\sub{0}]+\int d^4x
	g\sub{2}_{kl}\frac{\d g\sub{2}_{ij}[g\sub{0}]}{\d g\sub{0}_{kl}}.
\eeq
Therefore, there is a one-to-one correspondence between the terms in the
asymptotic expansion of holographic renormalization and our covariant expansion
in eigenfunctions of the dilatation operator. In particular, the non-local
terms in the two expansions are related, whereas the coefficients of the
logarithms - which are related to the conformal anomaly - are just proportional
to each other. This completes our demonstration of the equivalence of the two 
methods.

The new formulation, however, is advantageous over the standard 
method in that  the on-shell action is expressed entirely in terms of the extrinsic
curvature coefficients, for arbitrary $d$. Furthermore, the one-point function
in the presence of sources is also expressed simply in terms of one of the 
extrinsic curvature coefficients. 
The asymptotic analysis is done once, for all $d$, resulting
in generic recursion relations for the extrinsic curvature coefficients.
 
To complete the near boundary analysis one then just needs to solve the 
recursion relations for a given dimension $d$. The key ingredient in our
method which allows for these improvements is the functional relation
between the canonical momenta and the on-shell action, namely
\beq
\p^{ij}=-\frac{1}{2\k^2}\sqrt{\g}\left(K\g^{ij}-K^{ij}\right)=
	\frac{\d S_{\rm on-shell}}{\d\g_{ij}},
\eeq
or
\beq\label{HJ}
K\g^{ij}-K^{ij}=\frac{2}{\sqrt{\g}}\frac{\d}{\d\g_{ij}}
	\int_{\S_{r}}d^dx\sqrt{\g}(K-\l).
\eeq
Inserting the covariant expansions for $K^i_j$ and $\l$ we can relate
the coefficients of the on-shell action to those of the extrinsic curvature
as
\beq\label{functional}
\framebox[\width]{
\begin{minipage}{5.in}
\begin{eqnarray*}
K\sub{2n}^i_j=\l\sub{2n}\d^i_j-\frac{2}{\sqrt{\g}}\int d^dx\sqrt{\g}\g_{kj}
	\frac{\d}{\d \g_{ik}}\left(K\sub{2n}-\l\sub{2n}\right),\,\,\,0\leq n\leq 
	\frac d2,\\
\tilde{K}\sub{d}^i_j=\tilde{\l}\sub{d}\d^i_j-\frac{2}{\sqrt{\g}}\int d^dx\sqrt{\g}
	\g_{kj}\frac{\d}{\d \g_{ik}}\left(\tilde{K}\sub{d}-\tilde{\l}\sub{d}\right).
\end{eqnarray*}
\end{minipage}}
\eeq
The trace of these equations then gives
\beq\label{trace}
\left(1+\d_D\right)K\sub{2n}=\left(d+\d_D\right)\l\sub{2n},\,\,\,0\leq n\leq 
	\frac d2,\phantom{more space}\left(1+\d_D\right)\tilde{K}\sub{d}=
	\left(d+\d_D\right)\tilde{\l}\sub{d}.
\eeq
Since we know how the coefficients transform under the dilatation operator, 
these relations completely determine $\l$ in terms of the trace of the extrinsic
curvature. Namely we obtain the significant result
\beq\label{result_1}
\l\sub{2n}=\frac{(2n-1)}{(2n-d)}K\sub{2n},\,\,0\leq n\leq \frac d2-1,\qquad
	\tilde{\l}\sub{d}=\frac{d-1}{2}K\sub{d},\qquad
	\tilde{K}\sub{d}=0. 
\eeq
The coefficient $K\sub{2n}^i_j$ are only determined for $n<d/2$. If one does the 
computation for general $d$ then the corresponding expression has a first 
order pole at $d=2n$. A short computation using (\ref{functional}) 
shows that the residue of the pole is exactly $\tilde{K}\sub{d}^i_j$,
i.e.  the coefficient of the logarithmic term in $d$ dimensions,
\be
\tilde{K}\sub{d}^i_j = \lim_{n \to d/2} \left( (n- \frac{d}{2}) K\sub{2n}^i_j \right). 
\ee 
In practice one can also use this result in order to compute $K\sub{d-2}^i_j$ in  
$d$ dimensions from  $\tilde{K}\sub{d-2}^i_j$ in $d-2$ dimensions.

We thus arrive at a general closed form expression for
 the covariant counterterm action that renders the on-shell action finite:
\beq\label{result_5}
\framebox[\width]{
\begin{minipage}{5.in}
\begin{eqnarray*}
\,\,S_{ct}=\frac{(1-d)}{\k^2}\int_{\r=\e}d^dx\sqrt{\g}\left[
	\sum_{m=0}^{\frac d2-1}\frac{1}{(2m-d)}K\sub{2m}+\frac{1}{2}
	K\sub{d}\log\e\right] \\
\end{eqnarray*}
\end{minipage}}
\eeq
The rest of the analysis is now straightforward. First, by direct substitution
 of the covariant expansion of the extrinsic curvature into the first equation in 
(\ref{einsteinIIgr}) one finds a recursive relation for the traces of the
extrinsic curvature coefficients, namely
\beq\label{result_2}
\framebox[\width]{
\begin{minipage}{5.5in}
\begin{eqnarray*}
K\sub{2}=\frac{R}{2(d-1)},\phantom{more more more space here here}\\
	K\sub{2n}=\frac{1}{2(d-1)}\sum_{m=1}^{n-1}\left[K\sub{2m}_{ij}
	K\sub{2n-2m}^{ij}-K\sub{2m}K\sub{2n-2m}\right],
	\phantom{more}2\leq n\leq\frac d2\\
\end{eqnarray*}
\end{minipage}}
\eeq
Finally, inserting the values of $\l\sub{2n}$ and the 
traces of the extrinsic curvature we have determined in (\ref{result_1}) and
(\ref{result_2}) into the functional relation (\ref{functional}) one can
evaluate all coefficients recursively. In doing so, one sees that considerable
simplifications occur upon using the second equation in (\ref{einsteinIIgr}), 
which implies
\beq
\nabla_iK\sub{2n}^i_j-\nabla_jK\sub{2n}=0,\,\,\,0\leq n\leq\frac d2,\phantom{more}
	\nabla_i\tilde{K}\sub{d}^i_j-\nabla_j\tilde{K}\sub{d}=0.\\
\eeq
Note that although $K\sub{d}^i_j$ is non-local in general, its trace is local as follows
from (\ref{result_2}). Carrying out the above procedure is straightforward but 
the result becomes of forbidding complexity as one goes up in dimension.
The algorithm, however, could be implemented in a computer code
which would in principle calculate the counterterms and the holographic 
Weyl anomaly for any dimension. For illustrative purposes we quote the 
results for up to four dimensions,
\begin{eqnarray}\label{results}
\mathbf{d=2}&\NO \\
& K^i_j[\g]=\d^i_j+K\sub{2}^i_j+\ldots \NO \\ \NO \\
& K[\g]=d+P+\ldots \NO \\ \NO \\
& S_{\rm ct}=\frac{(d-1)}{\k^2}\int_{\r=\e}d^2x\sqrt{\g}\left[1-\frac 14
	 R\log\e\right]\NO \\ \NO \\
\mathbf{d=4} \NO & \\
& K^i_j[\g]=\d^i_j+P^i_j+\frac 12\left[\frac 12\left(P^{kl}P_{kl}-P^2\right)\d^i_j
	\right.\phantom{more more more space more}\NO \\&\left.\phantom{more space}
	-\frac{1}{(d-2)}\left(2R^i\phantom{}_{kjl}P^{kl}-PR^i_j+\square P^i_j
	-\nabla^i\nabla_jP\right)\right]\log\e+K\sub{4}^i_j+\ldots\NO \\ \NO  \\
& K[\g]=d+P+\frac{1}{2(d-1)}\left(P^{kl}P_{kl}-P^2\right)+\ldots\NO \\ \NO \\
& S_{\rm ct}=\frac{(d-1)}{\k^2}\int_{\r=\e}d^4x\sqrt{\g}\left[1+
	\frac{1}{(d-2)}P-\frac{1}{4(d-1)}\left(P^{kl}P_{kl}-P^2\right)
	\log\e\right]\NO \\ \NO
\end{eqnarray}
where we have introduced the sectional curvature tensor
\beq P_{ij}=\frac{1}{(d-2)}\left(R_{ij}-\frac{1}{2(d-1)}R\g_{ij}\right)\eeq
which transforms under Weyl rescalings of the metric $\d\g_{ij}=-2\g_{ij}\d\s$
as $\d P_{ij}=\nabla_i\nabla_j\d\s$. 

\subsection{Gravity coupled to scalars}

Having carried out in detail the near boundary analysis for pure AdS gravity
in our formalism, we will now briefly describe how the analysis can be
generalized to include scalars. In this case the matter action takes the form
\beq 
S_m=\int_Md^{d+1}x \sqrt{g} \left[\frac 12 g^{\m\n}\partial_\m\F_I \partial_\n
	\F_I+V(\F_I)\right].
\eeq   
Along with the gravitational field equations
(\ref{einsteinIIgf}) 
and equation (\ref{lambda}) for the on-shell action,
we now have the equations of motion for the scalar fields
\beq 
\ddot{\F}_I+K\dot{\F}_I+\square\F_I-\frac{\partial}{\partial\F_I}V(\F) =0 .
\eeq
In terms of the canonical momenta\footnote{Strictly speaking, the momenta are
densities and should include a factor of $\sqrt{\g}$ as we defined them earlier,
see for instance \cite{Wald}.  
Nevertheless, we will drop (with due care) this factor in this section as this
results in simpler equations.}
$\p^I=\dot{\F}_I$,
\beq 
\dot{\p}^I+K\p^I+\square\F_I-\frac{\partial}{\partial\F_I}V(\F) =0 .
\eeq
Next, we expand the canonical momenta and on-shell action in eigenfunctions of
the dilatation operator
\beq
\d_D=\int d^dx2\g_{ij}\frac{\d}{\d\g_{ij}}+
	\int d^dx(\D_I-d)\F_I\frac{\d}{\d\F_I}.	
\eeq
In addition to the expansions (\ref{cov_expansions}) we now have  expansions for
the canonical momenta of the scalar fields
\beq
\p^I[\g,\F]=\sum_{d-\D_I\leq s< \D_I} \p\sub{s}^I+\p\sub{\D_I}^I+\tilde{\p}
\sub{\D_I}^I\log{\rm e}^{-2r}+\cdots
\eeq
The crucial difference is that the momenta now depend on {\em all} bulk fields
and not just the induced metric. Moreover, depending on the scaling dimension
$\D_I$, the eigenvalues of the dilatation operator may not be integers anymore. 
The analysis is exactly analogous to that for pure AdS gravity, making essential
use of the functional relations (\ref{HJ}) and
\beq\label{HJ_scalar}
\p^I=-\frac{1}{\k^2}\frac{1}{\sqrt{\g}}\frac{\d}{\d\F_I}\int d^dx\sqrt{\g}
	\left(K-\l\right),
\eeq
which imply the key relation
\beq 
\left(1+\d_D\right)K+\k^2(\D_I-d)\p^I\F_I=\left(d+\d_D\right)\l.
\eeq
As for pure AdS gravity, this can be used to express the coefficients of
$\l$ in terms of those of the momenta. Then, inserting $\l$ into (\ref{HJ})
and (\ref{HJ_scalar}), the canonical momenta are determined iteratively.

As an illustration, consider the case of two scalar fields, $\F$ and $\S$, both of
scaling dimension $\D=3$ in $d=4$ and with potential that has a critical 
point at $\F=\S=0$. The most general potential compatible with these 
requirements is
\beq 
V(\F,\S)=\sum_{n=0}^\infty \sum_{m=0}^{n}\k^{n-2}V\sub{m,n-m}\F^m\S^{n-m}.
\eeq 
where $V_{(0,0)}=\L/\k^2$ is the cosmological constant, $V_{(0,1)}=V_{(1,0)}=0$,
i.e. there are no linear couplings, $V_{(1,1)}=0$ and $V_{(2,0)}=V_{(0,2)}=-3$,
i.e. the quadratic terms are diagonal in $\F$ and $\S$ and both have mass
$m^2=\D (\D-d)=-3$, and all other coupling $V_{(m,m-n)}$ are arbitrary.
The iterative approach determines the following on-shell action:
\beq\label{results_scalars}
\framebox[\width]{
\begin{minipage}{6.in}
\begin{eqnarray*}
S_{\rm on-shell}=-\frac{1}{\k^2}\int_{\r=\e}d^4x\sqrt{\g}\left\{
	\frac{d-1}{d-2}P-\frac 14\left[P_{ij}P^{ij}-P^2-\k^2\F(\square+P)\F
	\right.\right.\\\left.\left.-\k^2\S(\square+P)\S\right]\log\e+K\sub{4}-
	\l\sub{4}+\ldots\right\}-\int_{\r=\e}d^4x\sqrt{\g}W(\F,\S)\phantom{more space}
\end{eqnarray*}
\end{minipage}}
\eeq
where the ``superpotential'' is given by
\beq\label{results_superpotential}
\framebox[\width]{
\begin{minipage}{6.in}
\begin{eqnarray*}
W(\F,\S)=\frac{1}{\k^2}(d-1)+\frac 12\left(\F^2+\S^2\right)+\phantom{\hspace{2.in}}\\
	\frac{\k}{d-3}\left(V\sub{3,0}\F^3+V\sub{2,1}\F^2\S+V\sub{1,2}\F\S^2+
	V\sub{0,3}\S^3\right)+\\
	\k^2\left[\left(\frac 12 V\sub{4,0}-\frac{1}{4(d-3)^2}(9V\sub{3,0}^2+
	V\sub{1,2}^2)+\frac{d}{16(d-1)}\right)\F^4+\right.\\
	\left.\left(\frac 12 V\sub{3,1}-\frac{1}{(d-3)^2}(3V\sub{3,0}V\sub{2,1}+
	V\sub{1,2}V\sub{0,3})\right)\F^3\S+\right.\\\left.
	\left(\frac 12 V\sub{2,2}-\frac{1}{2(d-3)^2}(3V\sub{3,0}V\sub{1,2}+
	\phantom{more more space }\right.\right.\\\left.\left.
	3V\sub{0,3}V\sub{2,1}+2V\sub{2,1}^2+2V\sub{1,2}^2)+\frac{d}{8(d-1)}
	\right)\F^2\S^2+\right.\\\left.
	\left(\frac 12 V\sub{0,4}-\frac{1}{4(d-3)^2}
	(9V\sub{0,3}^2+V\sub{2,1}^2)+\frac{d}{16(d-1)}\right)\S^4+\right.\\
	\left.\left(\frac 12 V\sub{1,3}-\frac{1}{(d-3)^2}(3V\sub{0,3}V\sub{1,2}+
	V\sub{2,1}V\sub{3,0})\right)\F\S^3\right]\log\e +W\sub{4}+\ldots
\end{eqnarray*}
\end{minipage}}
\eeq
A direct computation shows that $W$ satisfies
\beq\label{superpotential}
V(\F_I)=\frac 12\left[\left(\frac{\partial W}{\partial\F_I}\right)^2
	-\frac{d\k^2}{d-1}W^2\right].
\eeq
and thus can be considered as a ``superpotential''. The AdS critical point 
of $V$ is also a critical point of $W$. This  together with 
(\ref{superpotential}) guarantee gravitational stability of the AdS 
critical point and of BPS domain-wall solutions of the gauged
supergravity \cite{Towns,SkenTowns}. The expression (\ref{superpotential}) can be 
derived in general in this formalism, i.e. for arbitrary dimension and scalar fields \cite{deBoer:1999xf}.
In this context one may also view (\ref{superpotential}) as a differential equation
that should be solved to determine the ``superpotential''. Since (\ref{superpotential})
is quadratic in $W$ it can only determine $W$ up to a sign. In contrast, our method 
{\em guarantees} that the ``superpotential'' satisfies (\ref{superpotential}), but
nevertheless there is no sign ambiguity in the determination of 
(\ref{results_superpotential}).
 
So far we have presented a method to solve the first order functional differential 
equations asymptotically. It would be, of course, desirable to solve these non-linear
equations completely for arbitrary sources, but this is beyond present capabilities. 
However, we can get an idea of the difficulty of the problem by considering the 
easiest part, namely the superpotential for one scalar field. The full superpotential
can be determined (up to sign that is fixed using the asymptotic solution for $W$)
by equation (\ref{superpotential}) which is a first order ODE, as
opposed to the first order functional differential equations that determine the rest
of the on-shell action. To solve for the superpotential we observe that 
it is possible to bring (\ref{superpotential}) into the
form of Abel's equation \cite{Kamke}:
\beq
y'(\psi)=\frac{v'}{v} y^3-y^2-\frac{v'}{v} y+1
\eeq
where $\psi=\sqrt{\frac{d\k^2}{d-1}}\f$,  $y={\rm coth}(u)$, $W=v {\rm cosh}(u)$,
and $v$ is related to the potential by $\frac{2(d-1)}{d\k^2}V=-v^2$. 
The general solution to
this equation is not known, but it can be solved in special cases. For example,
the potential
\beq
V(\psi)=-\frac{d(d-1)}{2\k^2}\cosh\left(\frac 23\psi\right)
\eeq
leads to a soluble equation, with solution
\beq
W(\psi)=\frac{(d-1)\cosh^{1/2}\left(\frac 23\psi\right)
\left(\cosh\left(\frac 23\psi\right)+\cosh^{1/2}(\g){\rm sech}^{1/2}\left(\frac 43\psi+\g\right)\right)}
	{\k^2 \sqrt{1+ \cosh(\g){\rm sech}\left(\frac 43\psi+\g\right)+ 
2 \cosh\left(\frac 23\psi\right) \cosh^{1/2}(\g){\rm sech}^{1/2}\left(\frac 43\psi+\g\right)}}
\eeq  
Here $\g$ is an arbitrary parameter, and the scaling dimension, $\D$, of the 
operator dual to the scalar field is $2d/3$. Both the potential and the
superpotential have $\f=0$ as a critical point. $W$ has an expansion around
zero
\beq
W(\psi)=\frac{d-1}{\k^2}\left(1+\frac 16\psi^2+\frac{1}{27}\tanh(\g)\psi^3+
\ldots\right)
\eeq
Since $\D=2d/3$ the term cubic in $\f$ has a dilatation eigenvalue 
$3(\D-d)=-d$ which is exactly the correct order where the asymptotic 
expansion of the on-shell action breaks down and an undetermined term arises.
In this case it happens that there is no logarithmic term, but a new 
parameter $\g$ appears at the correct order.

\section{Conclusions}

We reviewed in this paper how QFT data is encoded in geometry via the 
AdS/CFT correspondence and we developed a new, more powerful, calculational 
method. A central element in the extraction of the QFT data is the 
form of asymptotic solutions for Einstein and matter field equations.
The asymptotic solutions contain, after a number of terms that are uniquely 
determined in terms of the Dirichlet data, a term that is only 
partially determined by the asymptotic analysis. At the same order 
a logarithmic term may appear. We can now summarize the way the QFT
data are encoded as follows: 
\begin{itemize}
\item The undetermined coefficient encodes all correlation functions.
To uncover them one needs exact solutions with arbitrary Dirichlet data.
\item The relations that the undetermined coefficients satisfy
are related to QFT Ward identities, including anomalies.
\item The logarithmic term is also related to the conformal anomaly.
\end{itemize}
The conformal anomaly 
appears also as 
a coefficient of the logarithmic divergence of the on-shell action \cite{HS}. 
The relation between logarithmic divergences in the on-shell action 
and the logarithmic term in the asymptotic expansion of a (free) scalar
field has also recently appeared 
in the math literature \cite{GZ,FG2} 
in studies related to the so-called Q-curvature \footnote{
The Q-curvature is a generalization of the scalar curvature in
two dimensions: it satisfies analogous conformal transformation
properties.}. It would be interesting to understand the significance of 
the Q-curvature in terms of the dual quantum field theory. Let us also 
emphasise again that in terms of physics it is most important 
to obtain information about the undetermined coefficients 
by studying, for instance, the constraints implied by regularity in the interior.
 
We have presented a new method for obtaining renormalized correlation
functions, conformal anomalies and Ward identities of the boundary QFT from 
geometrical data. The method is a Hamiltonian version 
of the standard approach but with the radial coordinate playing the 
role of time. In this approach the renormalized one-point 
functions of the boundary QFT in the presence of sources are related to the canonical
momenta of the bulk fields. The near-boundary expansions of the standard method
are replaced by covariant expansions in eigenfunctions of the dilatation operator. 
This leads to  simple closed form expressions for the counterterms  
and for the renormalized one-point functions in terms of the coefficients
of the covariant expansion of the momenta, valid in all dimensions.
For the coefficients we derived 
general recursion relations (also valid in any dimension).
This leads to a more efficient algorithm for determining counterterms 
and one-point functions than in previous works. It would 
be important to integrate the recursion relation in general.
It seems likely that this would require a more intuitive and geometric
understanding of the conformal anomaly. 

\section*{Acknowledgments}

We would like to thank W. M\"{u}ck for comments on the first version
of this work. KS is supported by NWO.

\end{document}